\documentclass[pre]{revtex4}

\usepackage{amsmath}
\usepackage{graphicx}

\newcommand{\tP}{\tilde{P}}
\newcommand{\tW}{\tilde{W}}
\newcommand{\tN}{\tilde{N}}
\newcommand{\cW}{\mathcal{W}}
\newcommand{\twt}{\cW_t}
\newcommand{\ent}{\mathcal{S}}

\graphicspath{{./Figure/}}

\begin{document}

\author{A. Puglisi} \affiliation{CNR-ISC c/o Dipartimento di Fisica, Universit\`a
Sapienza, p.le A. Moro 2, 00185 Roma, Italy and Istituto Sistemi
Complessi (ISC), CNR, via dei Taurini 19 00185 Roma}

\author{S. Pigolotti} \affiliation{The Niels Bohr International
  Academy, The Niels Bohr Institute, Blegdamsvej 17, DK-2100
  Copenhagen, Denmark}

\author{L. Rondoni}
\affiliation{Dipartimento di Matematica and INFN, Politecnico di Torino, Corso
Duca degli Abruzzi 24, 10129 Torino, Italy}

\author{A. Vulpiani} \affiliation{Dipartimento di Fisica, CNR-ISC and
  INFN, Universit\`a Sapienza, p.le A. Moro 2, 00185 Roma, Italy}

\title{Entropy production and coarse-graining in Markov processes}

\begin{abstract}
  We study the large time fluctuations of entropy production in Markov
  processes. In particular, we consider the effect of a
  coarse-graining procedure which decimates {\em fast states} with
  respect to a given time threshold. Our results provide strong
  evidence that entropy production is not directly affected by this
  decimation, provided that it does not entirely remove loops carrying
  a net probability current. After the study of some examples of
  random walks on simple graphs, we apply our analysis to a network
  model for the kinesin cycle, which is an important biomolecular
  motor. A tentative general theory of these facts, based on
  Schnakenberg's network theory, is proposed.
\end{abstract}

\maketitle

{\bf To our friend and colleague Massimo
Falcioni, on his 60th birthday}

%%%%%%%%%%%%%%%%%%%%%%%%%%%%%%%%%
\section{Introduction}

The coarse-graining procedure is a fundamental ingredient of the
statistical description of physical
systems~\cite{M85,K00,CFLV08}.  By coarse-graining we
mean a procedure which reduces the number of observables to simplify
the physical description. For instance, it is used to describe the
behaviour of the physically relevant quantities, or {\em slow
variables}, which depends on the coupling among all variables
characterizing the system of interest, including the so-called {\em
fast variables}. The archetype of such a procedure is the treatment of
Brownian colloidal particles, immersed in a fluid, in terms of the
Langevin equation. In this sense any model meant to represent a
real phenomenon may be thought of as a coarse-grained, i.e.\ reduced,
description.  The purpose of a model is, indeed, to advance our
understanding of the object under investigation, by highlighting its
interesting features and discarding the irrelevant ones. In turn, the
roles of relevant and irrelevant characteristics depend on the purpose
of the analysis to be performed.  Furthermore, it isn't always obvious
which quantities should be listed as interesting, and which ones
should be neglected, especially if a new problem is to be tackled
\cite{M85,K00,CFLV08,AURIGA,DGRBC}.  Therefore, it is critical to
understand how specific physical observables depend on the
coarse-graining procedure.

%Classical examples of coarse-grained descriptions at different resolution levels include the steps meant 
%to connect the microscopic descriptions of systems of physical interest to the macroscopic ones. These 
%can be summarized as follows: \\
%%\begin{enumerate}
%1) microscopic deterministic level: $\Gamma$-space description (positions and
%momenta of the $N$ particles), Liouville equation\\
%2) microscopic stochastic level: $\mu$-space description (position and
%momentum of one particle), Boltzmann equation\\
%3) mesoscopic large scale level: $\mu$-space description by means of the Fokker-Planck equation\\
%4) macroscopic level: hydrodynamics, Fourier law, Navier-Stokes equations, etc.
%\end{enumerate}

Examples of coarse-grained descriptions at different resolution levels
include the steps meant to connect the microscopic descriptions of
systems of physical interest to the macroscopic ones, for instance the
passage from the deterministic $\Gamma$-space description (positions
and momenta of the $N$ particles) to the stochastic $\mu$-space
description (position and momentum of one particle), up to macroscopic
descriptions such as hydrodynamics, Fourier law, Navier-Stokes
equations, etc.

Other methods use the coarse-graining procedure in order to reduce the
number of variables, e.g.\ by a decimation method which suppresses the
fast variables, or perform a spatial coarse-graining, as in the
renormalization group approach. In these methods the coarse-graining
is parametrized by some threshold, here denoted as coarse-graining
level (CGL).  This paper is devoted to the investigation of the impact
of variations of CGL on the entropy production of non-equilibrium
systems.

In the last decades, the introduction of the so-called Fluctuation
Relations (FR) for deterministic dynamics, by Evans, Cohen, Morriss,
Gallavotti, Jarzynski and other authors brought about
important developements in the physics of far from equilibrium
systems~\cite{ECM,GC,CJ}. In the specific context of Markov processes, here discussed,
Lebowitz and Spohn~\cite{LS99} showed that the ``entropy production''
per unit time, measured on a time-interval $t$, $\cW_t$ say, is
described by a large deviation theory whose Cramer function, $C$,
enjoys the following symmetry property:
$$
C(\cW_t)-C(-\cW_t)=-\cW_t ~.
$$
This relation is the stochastic counterpart of the deterministic steady state FR, and we 
call it Lebowitz-Spohn FR, or simply FR.

We remark that the FR does not 
provide any specific information about the shape of $C$. 
Therefore, a coarse-graining procedure which preserves the Markovian character of the 
model should preserve the validity of the FR as well, although it may change the shape 
of the Cramer function. As a matter of fact, Rahav and Jarzynski~\cite{RJ07} argue that the validity 
of the FR is little affected by the coarse-graining procedure, even in nontrivial cases, 
such as those in which the decimation (or blocking) of variables results in the loss 
of the Markovian property. 

In the present paper, at variance with Ref.\cite{RJ07}, we do
not address the question of validity of the FR (which is always
satisfied by our models), but focus our attention on the effects of the decimation
procedures on the behaviour of the Cramer function.

Understanding how $C$ changes under variations of the CGL is relevant, e.g.\ to 
interpret experimental results, since they are always obtained at finite resolution 
(for instance in frequency, \cite{AURIGA}). Likewise, any model describing a real system
is necessarily affected by some degree of approximation or of idealism. For instance, 
the entropy production defined by Lebowitz and Spohn appears to be a rather abstract 
quantity, depending on the direct and inverse trajectories in the state space, as well 
as on their probabilities in the stationary state, cf.\ Section II. 
Furthermore, such a quantity cannot be measured in a direct way. Therefore, it needs to be
connected to directly measurable quantities, for its properties to be assessed.

\begin{figure}
\includegraphics[width=10cm,clip=true]{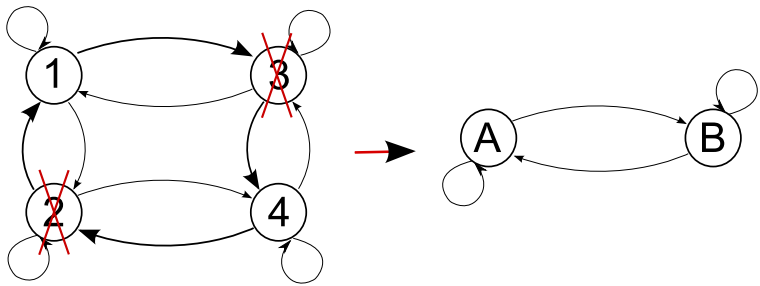}
\caption{A simple example of decimation which results in vanishing entropy
production. \label{fig:4states}}
\end{figure}

Naively, one may expect the entropy production computed through a model which encompasses 
lots of details of the system of interest to be higher than that computed through less
detailed models. Consider for example the Markov chain depicted in Fig.~\ref{fig:4states}a, 
with transition probabilities  $2 \to 1$ and $3 \to 4$ much larger than the remaining
ones. One may impose that a net current flows from $3$ to $1$ and from $2$ to $4$, 
by choosing transition probabilities $P_{3 \to 1} \ll P_{1 \to 3}$ and $P_{2 \to 4} \ll
P_{4 \to 2}$ and by tuning the other parameters so that all states have the same
stationary probability.
In this system, detailed balance does not hold, hence the mean entropy production is 
positive. However, the mean entropy production vanishes if the fast states $2$ and $3$
are decimated, and the Markov chain is reduced to the one represented in Fig.~\ref{fig:4states}b, 
where $A$ corresponds to the old state $1$ and $B$ to the old state $4$. Indeed, detailed 
balance, which implies a vanishing entropy production, holds in any Markov chain with two 
states. 

In the following we consider Markovian systems described by a
Master equation:
\begin{equation}\label{eq:master}
\frac{d P_n}{dt}=\sum_{l \neq n} P_l W_{l \to n} - W_n^0 P_n
\end{equation}
where $W_{l \to n}$ is the transition rate from state $l$ to state $n$, with $l,n\in[1,N]$, $N$ 
being the number of possible states of the process, $P_n(t)$ is the probability to stay in the 
state $n$ at time $t$, and 
\begin{equation}
W_n^0=\sum_{l \neq n} W_{n \to l}.
\end{equation}
We adopt the coarse graining procedure introduced in Ref.\cite{PV08} and described in 
Appendix~\ref{ap:dec}, which amounts to a decimation of all states whose characteristic 
times $\tau_n=1/W_n^0$ are smaller than a given $\Delta \tau > 0$. The resulting master 
equation for the surviving states may be written as 
\begin{equation}
\frac{d \tP_j}{dt}=\sum_{i \neq j} \tP_i \tW_{i \to j} - \tW_j^0 \tP_j
\end{equation}
with transition rates $\tW_{i \to j}$ as prescribed by \cite{PV08}, 
$i,j\in[1,\tN]$, $\tN<N$, and $\tW_j^0<1/\Delta \tau$ for all $j\in[1,\tN]$.  

Varying $\Delta \tau$ the number of the slow
states and the shape of the Cramer function $C_{\Delta \tau}(\cW_t)$
may in principle change.

In Section II, we present some numerical results for differently decimated Markov processes. 
In section III, we discuss the possibility of constructing a general theory, not yet available,
of the decimation effects. In Section IV, we draw some conclusions and discuss open problems. 
Appendix A illustrates the decimation procedure; Appendix B reports analytical results about 
the effect of decimation on the current in a single loop; Appendix C recalls the graph analysis 
of currents in a Markov system, based on Schnakenberg's theory; Appendix D describes in detail
the kinesine model mentioned at the end of section II; Appendix E lists the
main symbols used in the text.

%%%%%%%%%%%%%%%%%%%%%%%%%%%%%%%%%
\section{Some numerical results}

%%%%%%%%%%%%%%%%%%%%%%%%%%%%%%%%%%%%%%%%%%%%%%%%%
\subsection{Entropy production on a trajectory}

While the concept of entropy production, or energy dissipation, dates a long time back~\cite{DEGM}, 
only recently have the {\em fluctuations} of entropy production attracted 
particular interest, thanks to important theoretical and numerical results, supported by some 
experimental evidence. 
%Fluctuations are due to different
%realizations of the evolution of the system, which are equivalent in
%ergodic systems to consider different time intervals of a very long
%trajectory. 
For a trajectory of duration $t$ of a continuous time Markov process,
in which $m$ transitions $\omega_0 \to \omega_1 ... \omega_{m-1} \to \omega_m$ 
are observed, $\omega_i$ being
the $i$-th visited state, the following definition of entropy production has been 
given by Lebowitz and Spohn~\cite{LS99}: 
\begin{equation} \label{ls_def} \twt=\frac{1}{t}\ln \frac{W_{\omega_0
      \to \omega_1}W_{\omega_1 \to \omega_2}...W_{\omega_{m-1} \to
      \omega_m}}{W_{\omega_{1} \to \omega_0}W_{\omega_2 \to
      \omega_1}...W_{\omega_m \to \omega_{m-1}}}.
\end{equation}
It can be shown that the times $t_i$ at which the transitions occur affect the
numerical value of $\twt$ only with corrections of order $\mathcal{O}(1/t)$, 
negligible in the $t \to \infty$ limit. In the present paper, we 
consider this entropy production, introduced by Lebowitz and Spohn. Clearly, this 
quantity does not need to represent any real thermodynamic observable, since it
can be defined independently of the physical relevance of the Markov process at hand.
Nevertheless, as commonly done in the literature, we will refer to it merely as to 
``entropy  production''.

Even if in this paper we consider the case of continuous time, in the discrete
time case (Markov chains) one can use the same
definition~\eqref{ls_def}, by replacing $W_{i \to j}$ with the probability of a transition 
in a time step $\delta t$,  $P_{i\to  j}$. The relation between continuous and discrete time
quantities is incorporated in the equalities $P_{i \to j}=W_{i \to j}\delta t$ for $i
\neq j$, and $P_{i \to i}=1-W_i^0\delta t$.

The connections of $\twt$ with other definitions of entropy production rate are discussed in
Appendix~\ref{ap:graph}. Here, it suffices to recall that $\langle \twt \rangle=0$ in 
the steady state, if the invariant probability $P^{inv}_{\omega}$ of the process 
satisfies the detailed balance condition
\begin{equation} \label{detbal} \frac{W_{\omega \to
      \omega'}}{W_{\omega' \to
      \omega}}=\frac{P^{inv}_{\omega'}}{P^{inv}_{\omega}}.
\end{equation}
The system is in equilibrium if eq.~\eqref{detbal} holds.  
If detailed balance does not hold, one has $\langle \twt \rangle >0$. 
Let $C$ be the Cramer function of the probability density function (pdf) $f$ of 
$\twt$, in the steady state, i.e.\ let $C$ be defined by
\begin{equation}
C(\twt)=-\lim_{t \to \infty} \frac{1}{t} \log[f(\twt)].
\end{equation}

In numerical calculations, the Cramer function $C$ must be
approximated by its finite time counterparts, $C(\twt) \approx
-\log[f(\twt)]/t$. Therefore, in our calculations, we have chosen
times $t$ large enough that further growths of the averaging times
practically do not affect our results.

Lebowitz and Spohn have shown that the condition 
\begin{equation} \label{fr}
C(\cW_t)-C(-\cW_t)=-\cW_t,
\end{equation}
is better and better approximated as the time $t$ grows~\cite{LS99}.
The $t \to \infty$ limit of relation~\eqref{fr} is known as a Steady State Fluctuation Relation (SSFR). 
It does not provide the shape of $C$, but only a symmetry property of $C$. 
Remarkably, $C$ is system-dependent~\cite{BPRV08}, while (\ref{fr}) holds quite in general. 

In the following, we address the question of the dependence of $C$ on the CGL which, in the 
decimation procedure of \cite{PV08}, is parametrized by the threshold time $\Delta t$. 
The protocol of Ref.\cite{PV08}, eliminates all states $i$ with 
average exit time $\tau_i<\Delta t$, and requires the surviving states to have 
re-normalized transition rates $\tW_{\omega \to \omega'}$. 
Denote by $f_{\Delta t}$ and by $C_{\Delta t}$ the pdf of the entropy production
and its Cramer function, for the decimated process with threshold time $\Delta t$. 

The present investigation suggests the conjecture that the entropy production does 
not depend sensibly on the precise properties of fast and slow variables of the Markov process: 
it only depends on the currents flowing in the system.

%%%%%%%%%%%%%%%%%%%%%%%%%%%%%%%%%%%%%%%%%%%%%%%%%%%%%%%%%%%%%%%%%%%
\subsection{Results on $1d$ and $2d$ regular lattices}

%Any continuous time Markov process can be considered as a weakly
%correlated continuous time random walk on a graph: correlation is
%short, in fact jump probabilities depend only on the current state. 

Let us begin focusing on continuous time random walks on simple
topologies, i.e.\ on regular lattices with periodic boundary conditions,
with random transition rates restricted to nearest neighbours. To simplify 
the procedure, we require every state $n$, $n=1,...,N$, to have characteristic 
(exit) time $\tau_n=1/\sum_{l \neq n} W_{n \to l}$, belonging to a set $\{\tau^{(1)},...,\tau^{(M)}\}$ 
such that $\tau^{(\alpha-1)} \ll \tau^{(\alpha)} \ll \tau^{(\alpha+1)}$. 
This condition corresponds to a separation of time-scales 
which represents a mild requirement for the decimation protocol of \cite{PV08}
to apply. 
Transition rates may then be chosen to have, or not to have, a preferential direction, 
in order to allow, or to prevent, a positive entropy production. For instance, 
entropy production can be positive in $1d$ lattices, only if some rates obey 
$W_{i \to i+1}/W_{i \to i-1} \neq 1$  (cf.\ Appendix~\ref{ap:graph}, for a more precise 
condition, based on the notion of affinities).

\begin{figure}[ht]
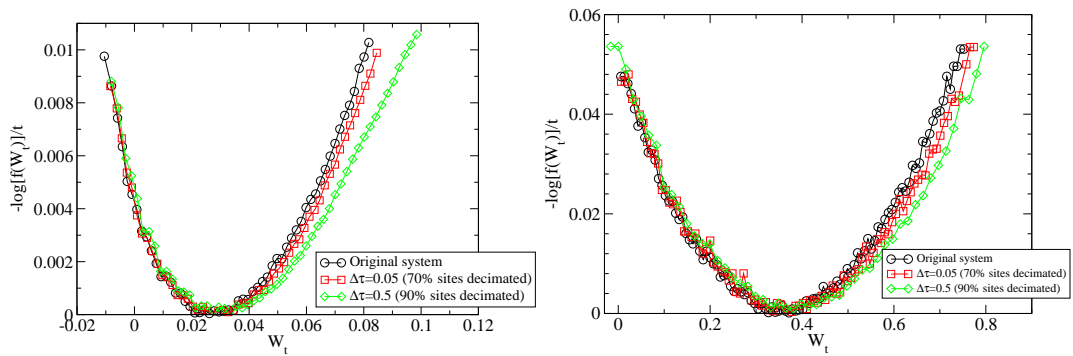

\includegraphics[width=7cm,clip=true]{pdf_1d_seitipi.eps}
\includegraphics[width=7cm,clip=true]{pdf_1d_tretipi.eps}
\caption{Numerically computed Cramer function for entropy production
  distribution in $1D$ continuous time random walks with p.b.c. In 
  each plot a comparison among different coarse graining levels is
  shown. In both panels, each state of the non-decimated
  system may take one out of three possible average exit times:
  $1$ ($10\%$ of states), $0.1$ ($20\%$ of states) and $0.01$ ($70\%$
  of states). In the left panel, the probability of jumping to the right is $0.4$ ($60\%$ of
  states) or $0.6$ ($40\%$ of states). In the right panel, the probability of jumping
  to the right is $0.4$. In the left frame we have $N=100$ and $t=2\cdot 10^3$. In the right frame we have
  $N=300$ and $t=10^3$. The numerical computation has been performed
  with the Gillespie algorithm~\cite{gill}, where the actual
  probability of a transition is the product of the transition rate by
  the characteristic time. \label{fig:1d}}
\end{figure}

Simulations for regular lattices show a striking robustness of
the entropy production Cramer function with respect to decimation. The
numerically computed Cramer function $-\log f_{\Delta t}(\twt)/t$ for $1d$ chains is
plotted in Figure~\ref{fig:1d} for different CGL.  The figure shows
that decimating $90\%$ of the system, i.e. leaving only the slowest
states, the fluctuations of entropy production remain substantially
the same. The result does not seem to depend on the details of the
transition rates, but only on the separation of time-scales.

In the following sections, we show that the entropy production is 
not directly related to the properties of fast and slow states {\em per se} either, while it
seems reasonable to conjecture that it is closely related to the {\em currents}
flowing in the system. These are global quantities, rather than local ones, which
depend on the topology and on the interplay among all transition rates, an idea that
may be understood in simple terms,
as follows. In the case of a random walk on a ring (a 1-dimensional lattice with 
periodic boundary conditions), one may write
\begin{equation}
  \frac{W_{\omega_0 \to \omega_1}W_{\omega_1 \to \omega_2}...W_{\omega_{n-1} \to \omega_n}}{W_{\omega_{1} \to \omega_0}W_{\omega_2 \to \omega_1}...W_{\omega_n \to \omega_{n-1}}}=\left(\frac{W_{forw}}{W_{back}}\right)^m\times \mathcal{R}
\end{equation}
where $W_{forw}=W_{1 \to 2}W_{2 \to 3}...W_{N-1 \to N}W_{N \to 1}$,
$W_{back}=W_{1 \to N}W_{N \to N-1}...W_{3 \to 2}W_{2 \to 1}$, $m$ is
an integer and $\mathcal{R}\approx\mathcal{O}(1)$ is a correction
term. In presence of a mean current $J>0$, one has $m=[G(t)
t]$ where $G(t)$ is the current computed in the time window $(0,t)$
(i.e. $\lim_{t \to \infty} \langle G(t) \rangle=J$) and $[ a ]$ indicates
the integer part of $a$. This leads to the relation
\begin{equation}
\twt\approx G(t) \log\frac{W_{forw}}{W_{back}}+\mathcal{O}(1/t).
\end{equation}
Since the decimation protocol eliminates fast states and modifies transition
rates in order to leave basically unaltered the currents connecting the
surviving states, the fluctuations of entropy
production should not be sensibly affected by the decimation procedure. Observe, however,
that this current conservation is only approximate, not exact. The modification of the
current produced by the decimation can indeed be estimated analytically in simple cases,
as in systems whose states form a single loop. In Appendix \ref{ap:singleloop}, we argue 
that the correction should be generally small and related to the ratio of the times spent in the 
fast states and in the slow ones.

\begin{figure}[ht]
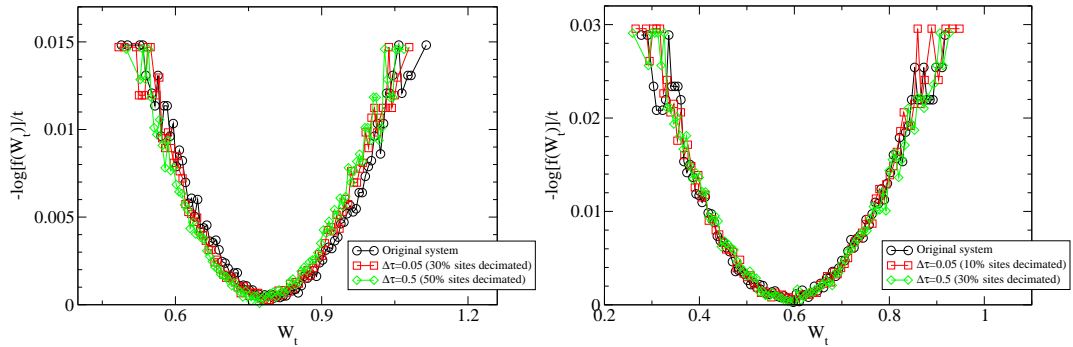

\includegraphics[width=7cm,clip=true]{pdf_2d_meglio.eps}
\includegraphics[width=7cm,clip=true]{pdf_2d.eps}
\caption{Approximated Cramer function for entropy production
  distribution in continuous time random walks on a $2D$ squared
  lattice (nearest neighbours) with p.b.c. In each plot a comparison
  among different coarse graining levels is shown. In both cases the
  probabilities of jumping to one of the four nearest neighbours are
  biased to give a net current in one direction. The left and the right panels
differ by the values of the exit times. Left: the states
  have exit times $1$ ($50\%$ of states), $0.1$ ($20\%$
  of states) and $0.01$ ($30\%$ of states). Right: the states have
  exit times $1$ ($70\%$ of states), $0.1$ ($20\%$ of states) and
  $0.01$ ($10\%$ of states). In both cases $N=100$ and $t=1000$ \label{fig:2d}}
\end{figure}

Similar results are reported in Figure~\ref{fig:2d} for 2D regular
square lattices, where jumps occur among nearest neighbours: even with
this topology, the fluctuations of entropy production appear not to be
affected by the CGL, although the result is not as robust as in the $1D$ case. 
Indeed, a substantial change in entropy production can be observed if the system 
contains a very large number of fast states to be decimated. Nevertheless, it still 
is interesting to realize that the Cramer function has not changed substantially,
even after $50\%$ of the original system has been decimated. 
Note that the square lattice topology
is drastically altered by decimation: states which have not been
decimated remain connected by chains of transitions, but the system is not planar 
nor a regular lattice anymore. Currents in this case may still be defined within 
a more general graph theory~\cite{S76}, such as that discussed in Appendix~\ref{ap:graph}.

%%%%%%%%%%%%%%%%%%%%%%%%%%%%%%%%%%%%%%%%%%%%%%%%%%%%%%%%%%%%
\subsection{Results on graphs with fast and slow loops}

\begin{figure}[ht]
\includegraphics[width=7cm,clip=true]{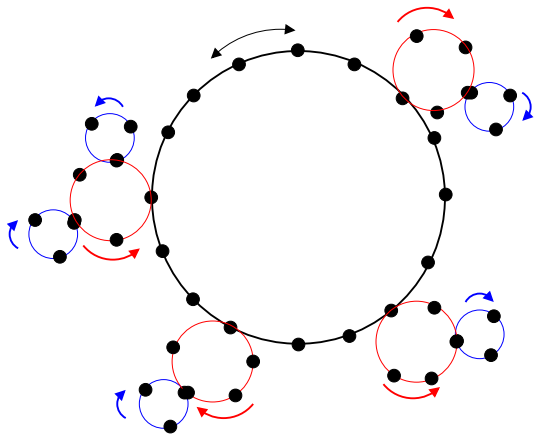}
\caption{Sketch of a graph made of three levels of nested loops: in
this example, the main level loop  has no preferential
direction, while the second and third levels are made of smaller
loops with faster states and preferential directions.
  \label{fig:topi}}
\end{figure}

Guided by the conjecture that the fundamental ingredient for entropy
production is the current flowing in a circuit, we construct Markov
processes composed of independent loops joined by a single interchange
state. The general structure of this graph is illustrated in
Fig.~\ref{fig:topi}. The main slow loop is decorated by fast loops
(first level), which are on their turn decorated by faster loops
(second level), etc. After decimation, one may encounter different
situations:

\begin{enumerate}
\item the new and the old structures have the same topology,
i.e.\ only pieces of loops have been suppressed but the number and
position of loops is the same;
\item all loops of the faster (outer) level are suppressed;
\item all loops of the two fastest levels are suppressed;
\item and so on;
\end{enumerate}
Loops at the same level have similar properties and, in particular, are
chosen to have, or not to have, a positive entropy production, i.e. to have or
not to have a preferential direction in their transition rates.

\begin{figure}[ht]
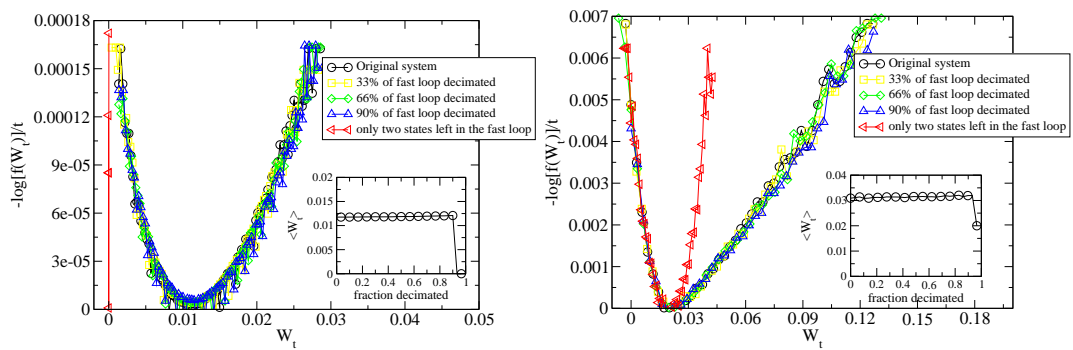

\includegraphics[width=7cm,clip=true]{gra_progressive2.eps}
\includegraphics[width=7cm,clip=true]{gra_progressive.eps}
\caption{Approximate Cramer function for entropy production
  distribution in continuous time random walks on a graph similar to that of
  Figure~\ref{fig:topi}, with two hierarchical levels. A comparison
  among different coarse graining levels is shown. The main loop is
  made of $100$ states with average exit time $1$ and preferential
  direction given by balanced (left) or unbalanced (right) transition rates. In the case of unbalanced rates, they are $0.6$ toward left and  $0.4$ toward right. The second level loops have $30$ states and a
  bias in the transition probabilities ($0.8$ vs. $0.2$) chosen to
  give a preferential direction, while their characteristic times
  range from $0.1$ to $0.7$. In all simulations $t=1000$.
  \label{fig:gra1}}
\end{figure}

\begin{figure}[ht]
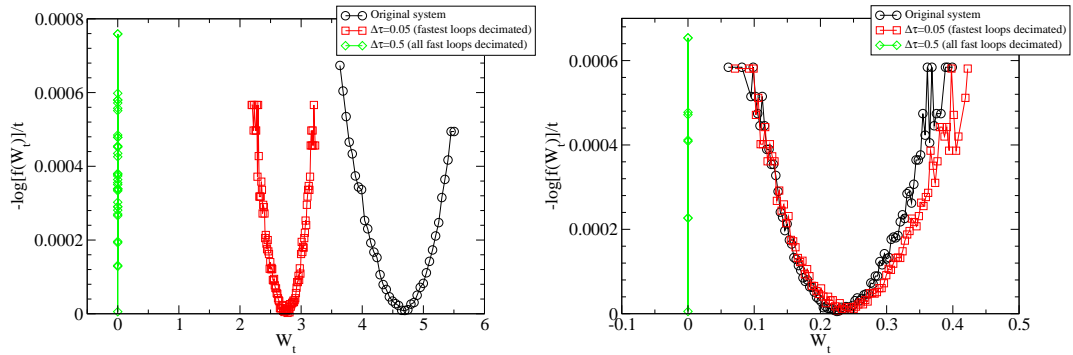

\includegraphics[width=7cm,clip=true]{pdf_gra2.eps}
\includegraphics[width=7cm,clip=true]{pdf_gra.eps}
\caption{Approximate Cramer function for entropy production
  distribution in continuous time random walks on the topologies of
  Figure~\ref{fig:topi}, with three hierarchical levels of nested
  loops and $t=10^4$. In each plot a comparison among different coarse grained
  levels is shown. The main loop is made of $100$ states with average
  exit time $1$ and no preferred direction. The second level loops
  (each with $10$ states) have average exit time $0.1$ and a bias in
  the transition probability chosen to give a preferential
  direction. The third level loops (each with $5$ states) have average
  exit time $0.01$.  The difference between the left and right frames
  is in the transition rates of this third level. Left: third level
  loops have a preferential direction. Right: third level loops without
  preferential direction.
 \label{fig:gra2}}
\end{figure}

The computed Cramer function for the entropy production of the case of Figure~\ref{fig:topi}
is reported in Figures~\ref{fig:gra1} and~\ref{fig:gra2}. In Figure~\ref{fig:gra1}
the states of the fast loop have slightly different characteristic
times, allowing a progressive decimation of the fast loop. At a
decimation threshold such that the fast loop is still alive, even if
made of only three states (blue curve), we are in situation 1 and the
Cramer function of the entropy production is very close to that of the
non-decimated system (black curve). A further increase of the
decimation threshold makes the fast loop disappear: it remains with
only two states and the system falls in situation 2, where the Cramer
function of the entropy production has a sudden macroscopic change
(red curve). The inset of Figure~\ref{fig:gra1} shows
$\langle \cW_t \rangle$ as a function of the decimated percentage of the fast
loop: neglecting a very weak growth, $\langle \cW_t \rangle$ appears practically 
constant, until the fast
loop is not reduced to a $2$-states branch. If the main loop is
configured to have a non-zero current, this is what remains at that
point, otherwise the fluctuations of $\twt$ are reduced to a very
narrow and symmetric peak around zero.

Figure~\ref{fig:gra2} shows other cases with three levels of
loops. The Cramer function of the entropy production always changes
when a level of current-carrying loops is entirely removed.

%%%%%%%%%%%%%%%%%%%%%%%%%%%%%%%%%%%%%%%%%%%%%
\subsection{A model from molecular biology: 
coarse graining of the Kinesin's network}

\label{sec:kin}

The examples in the previous sections suggest that the entropy
production is weakly affected by the coarse graining procedure, apart
from the cases in which loops contributing significantly to the
entropy are destroyed and the entropy production undergoes an abrupt
decrease.  A natural question is whether this phenomenology is a
peculiarity of the model introduced here, or similar
behaviors pertain to other realistic models of non-equilibrium 
systems.  Biochemical reactions are often characterized by non-equilibrium 
processes acting over different timescales and thus afford an
ideal benchmark for the ideas proposed here. In this subsection, we
study the effect of coarse graining on a recent network model of the
kinesin motor cycle \cite{LL07}.

Kinesins are a common category of motor proteins \cite{H01} that are
used for transport on microtubules in eucaryotic cells. Like many other
non-equilibrium reactions inside cells, kinesin is
powered by ATP. A number of experiments during the last decades
elucidated many structural and dynamical details of this systems. In
particular, it is now understood that kinesins are made of two
identical heads, that walk on microtubules with a ``hand-over-hand''
mechanism \cite{YTVS04}, alternating their position at the front.

\begin{figure}[htb]
\includegraphics[width=6cm]{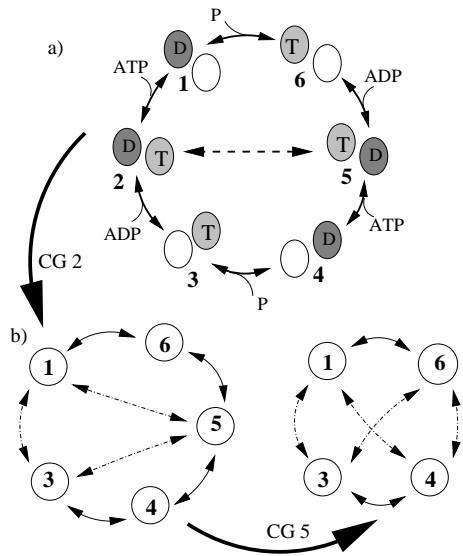}
\caption{Scheme of the transition network of the kinesin model
  \cite{LL07}. a) Reaction network. The states, numbered
  from $1$ to $6$, are characterized by the two heads bound to ATP
  (A), ADP (D) or free. The molecules bound (or released) during the
  chemical transitions are shown as connected to the arrows. The
  dashed arrow represents the mechanical transition, in which kinesin
  makes its step on the microtubule. b) Kinesin network after coarse
  graining the fastest states, first state $2$, then state
  $5$. Dot-dashed arrows represent the new transitions appearing as a
  consequence of the coarse graining procedure.}
\label{kinesin_scheme}
\end{figure}

The model proposed in \cite{LL07} describes both the
ATP-driven chemical reactions and the mechanical step in which the two
heads swap. The multiple cycle structure of the reaction is given by
this chemomechanical nature and by considering the fact that ATP may
be burned by both heads. The scheme of the reaction and the possible
transitions are illustrated in Fig.(\ref{kinesin_scheme}).
Each of the two heads may be in three different configurations: free,
bound to ATP and bound to ADP, resulting in $3\times3=9$ possible
states. However, the motor is believed to work ``out of phase'',
i.e. states in which the two heads are in the same configuration are
unlikely to be observed. This reduces the model to the $6$ states
represented in the scheme of Fig. (\ref{kinesin_scheme}). Of all the
possible transition among the states, only those which are consistent
with experimental observations are considered in the model and shown
in the diagram. Clearly, the assumptions above (in particular that of
considering only $6$ states) already imply some level of coarse
graining with respect to the complete  problem. However,
the effect of these assumptions on the entropy production is hard to
determine, since it would be difficult to construct a more detailed
model, from the available experimental results. We then take the model 
of Fig. (\ref{kinesin_scheme}) as our starting point,
and study the effect of decimating the states of the system.

The transition rates of the model depend on the ADP, ATP and P
concentration and on the load force $F$ of the molecular
motor. Moreover, the parameters determining these rates have a slight
dependence on the kind of the experiment one wants to reproduce, since
different experiments work in different conditions and may use
different kinds of proteins in the kinesin family. We determined the
rates by choosing the parameters fitting the experiment of Ref.\cite{CC05}
and assumed fixed concentrations of
$[ADP]=[ATP]=[P]=1\mu M$ (micromoles) for simplicity. We then consider two different
cases: one without work load and one with a work force equal to
$F\approx 5 pN$ (piconewton). Details on the derivation of the rates
and numerical values are given in Appendix (\ref{app_kin}).

\begin{figure}[htb]
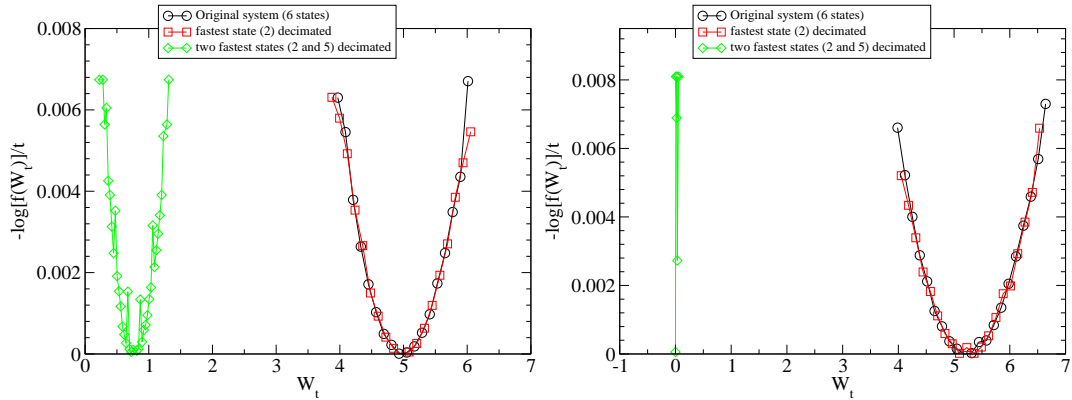

\includegraphics[width=7cm,clip=true]{pdfload.eps}
\includegraphics[width=7cm,clip=true]{pdfnoload.eps}
\caption{Pdf's of the entropy production per unit of time. Black curve
  is for the original model with 6 states. Red is after decimation
  of state 2. Green is after decimation of states 2 and 5. Left panel: the
  case without load. Right panel: the case with load.}\label{fig:kin}
\end{figure}

In both cases (with and without load), state $2$ is the fastest and
state $5$ is the second fastest. We compare then the entropy
production of the complete model, of the model in which state $2$ has
been adiabatically eliminated and of the model in which both $2$ and
$5$ have been eliminated. The pdf of finite-time-averaged entropy
production $W_{t_{max}}/t_{max}$, obtained from $10000$ realizations of
trajectories of length $t_{max}$ is shown in Figure~\ref{fig:kin},
without load in the left panel, with load in the right panel. Decimation of state 2 leaves the
pdf unchanged. In the case without load, further decimation of state 5 
changes abruptly the pdf to a close-to-zero peaked pdf.

This model reproduces the same scenario of the ``fast loop-slow
loop'' model of the previous section. The first coarse graining
strongly alters the structure of the network, but its effect on the
entropy production and its fluctuations is barely
noticeable. Conversely, decimating one more state drastically reduces
the entropy production. Notice also that the most irreversible
transition in the original model is the ``mechanical'' transition between
state $2$ and $5$, since $W_{25}/W_{52}=3\ 10^3$ for the
load-free case and $\approx 57$ for the loaded case. In the sense
specified in the next section, the information about the
irreversibility of this transition is lost, when the level of coarse
graining is too large.

%%%%%%%%%%%%%%%%%%%%%%%%%%%%%%%%%
\section{Tentative theory}

Consider a continuous time Markov process: each state $n$ can
be seen as a vertex of a graph, and transitions $n\to n'$ with a
positive rate correspond to edges (also called links) between $n$ and
$n'$. As in \cite{LS99}, we assume that the transition $n \to n'$ has 
a positive rate whenever the inverse transition $n' \to n$ does. As illustrated in 
Section 2.B, the entropy production of random walks on 1D rings is closely
related to the current flowing in the ring. In this Section, we
attempt to generalize this observation to generic
graphs~\cite{bc05}. The main tool for this purpose is a decomposition
in fundamental cycles, which is illustrated in
Appendix~\ref{ap:graph}. In the example of Figure~\ref{fig:topi}, the
fundamental cycles are nothing but the loops.

Let us introduce a different functional $Q_t$:
\begin{equation}
Q_t=\sum_\alpha A(\vec{C}_\alpha)G_\alpha(t)
\end{equation}
which depends only on a
few ``structural properties'' of the process: the {\em fundamental cycles} $\vec{C}_\alpha$
(i.e.\ a property of the graph), their {\em affinities} $A(\vec{C}_\alpha)$ and their
fluctuating {\em currents} $G_\alpha$, averaged over a time interval $(0,t)$, which depend
also on the transition rates.

\begin{figure}[ht]
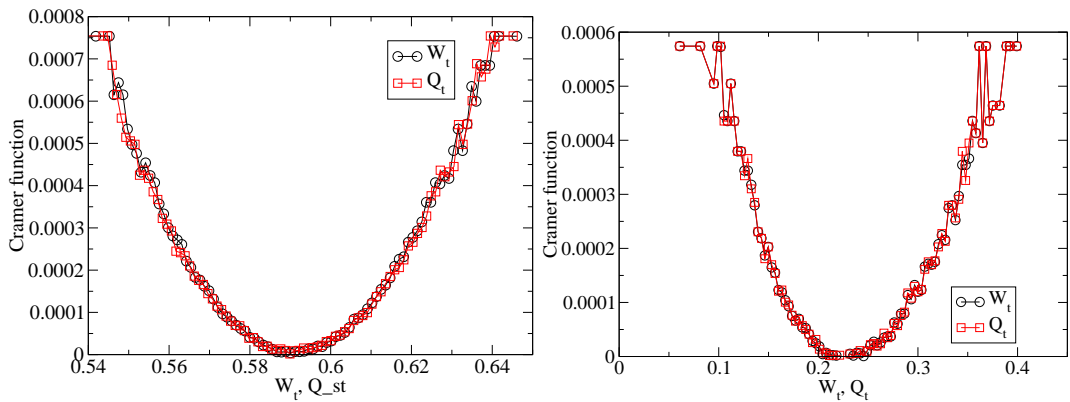

\includegraphics[width=7cm,clip=true]{w_q_2d.eps}
\includegraphics[width=7cm,clip=true]{w_q_ring.eps}
\caption{Equivalence of the Cramer function for $\cW_t$ and
$Q_t$ at $t=10^4$.  Left: random walk on a 2d lattice with same parameters as in
Figure~\ref{fig:2d} (right). Right: random walk on a nested loop graph
with same parameters as that in Figure~\ref{fig:gra2} (right).
\label{fig:equiv}}
\end{figure}

In various numerical simulations, we have verified that the fluctuations of $Q_t$ 
and those of $\twt$ are practically indistinguishable, at large times: 
cf.\ Figure~\ref{fig:equiv} for two
examples. It is also known that $\langle Q_t \rangle
= \langle \twt \rangle$ for large times~\cite{AG07}.

To obtain a complete theory, it remains to show that the ``structural
properties'' are not affected by the decimation; this task can be subdivided
in three steps:

\begin{enumerate}

\item examine the fate of fundamental cycles after decimation of one fast state;
there are three possibilities: {\bf i.} cycles may be destroyed, {\bf ii.} transformed into different
ones, {\bf iii.} new cycles may be created;

\item derive the corresponding variation of affinities;

\item obtain the values of currents after decimation.

\end{enumerate}

Concerning task
$2$, it is easy to realize that affinities do not change in
transformed cycles, while new cycles have zero affinity, so they do not
contribute to the entropy production. Disappearing cycles pose, instead, 
a difficult question: numerical simulations show that they are usually small 
and that the affinity lost with their removal is equally
small. At the moment, however, we do not have an analytical estimate of this
quantity. Task 3 is a hard problem too: the stationary value of currents must satisfy
many coupled Kirchhoff equations and depends on the properties of the
whole graph. Numerical simulations suggest that average currents are
not drastically influenced by our decimation procedure. One rough explanation
of this fact can be given for a system $\Sigma$ with small entropy production, obtained 
from a perturbation of an equilibrium system $\Sigma^0$. Indeed, one may assume a 
linear relation between the affinities $A$ and the average currents 
$J_\beta$ of $\Sigma$, of the form $A(\vec{C}_\alpha)=\sum_{\beta} L_{\alpha\beta} J_\beta$,
with coefficients $L_{\alpha\beta}$ determined only by the properties of
$\Sigma^0$. If the decimation procedure, which replaces $\Sigma$ with a new system $\Sigma'$,
leaves substantially unaltered the invariant probability of the surviving states, it
is reasonable to assume that the decimated system $\Sigma'$ is another small
perturbation of $\Sigma^0$. Then, the linear relation between affinities
and currents of $\Sigma'$ retains the same coefficients $L_{\alpha\beta}$,
leading to the conclusion that the currents are conserved under decimation, 
if affinities are.

We now discuss the consequences of decimation on cycles and
their affinities.
%%%%%%%%%%%%%%%%%%%%%%%%%%%%%%%%%%%%%
%\subsection{``Quasi-isomorphism'' of decimated graph}
A maximal (also called ``spanning'') tree $T$ is found on the original
graph $G$. This tree includes all $N$ vertices (states) and only a
part ($N-1$) of the original number $E$ of edges. All pairs of
vertices are connected by a unique path on this tree. All edges left
out from the tree (a number $\nu=E-(N-1)$) are called ``chords''. A 
chord connecting vertices $i$ and $j$, attached to the unique
path connecting $i$ and $j$ along the tree, is a closed loop. All
loops generated in this way constitute the set of fundamental loops,
which become ``cycles'' when orientation is taken into account. These
fundamental cycles determine the statistics of $Q_t$ and therefore of
$\twt$, cf.\ Appendix~\ref{ap:graph} for the details.
In the cases discussed below, the removal of a
vertex using our decimation procedure preserves almost exactly the fundamental cycles 
and their affinities; the small variations  observed are
due to the possibile reduction of whole $3$-loops to $2$-loops
(i.e.\ simple links) corresponding to a total loss of the affinity of
the original $3$-loop. The impact of this unfortunate event is
difficult to estimate, because it depends on the topology of the
graph: the removal of a vertex may lead to a crunch of a number of loops
smaller than or equal to the degree of the removed vertex. The amount
of lost affinity for each reduced loop is expected to be small, since it is
associated with a small loop, and correspondingly small should be the loss
in current and in entropy production. It is remarkable that the exit times of 
states do not affect the affinities, although they can affect the currents. 

Nevertheless, decimation may affect the large loops as well; a progressive 
and repeated removal of vertices may eventually reduce a large loop to a 
$2$-loop. Unfortunately, controlling these events goes beyond our mathematical ability, 
therefore, the size of the error in the conservation of fundamental cycles 
under decimation remains an open question. 

In the figures, all black objects (vertices and links) are related to
the original graph, red objects are the new ones
formed after decimation. Solid links are part of the maximal tree,
dashed links are chords. When the state labelled by $0$ is removed, it is
linked to some other states collectively denoted as $j$: $j$-states
(linked to $0$) are in number of $n$. These links are broken and all
pairs of states $j$ and $j'$ (previously connected to $0$) are
connected among each other with a new transition rate
$W_{j->0}W_{0->j'}/W_0^0$ (or, if the link $j-j'$ already exists, its
transition rate is updated adding that amount). We consider the
simplest case where one chord at most is involved in the decimation
procedure. With this assumption, three possibilities can be encountered:

\begin{figure}[ht]
\includegraphics[width=6cm,clip=true]{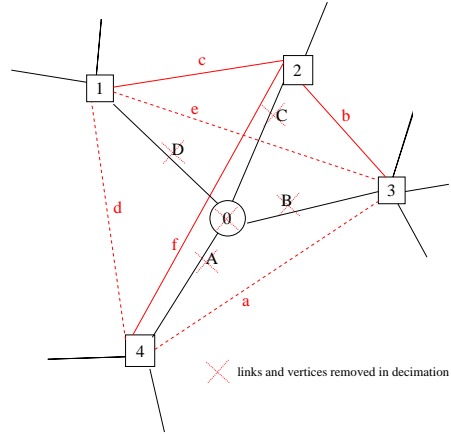}
\caption{An example of state-removal where no chords belong to the
  subgraph involving the removed state. Black objects (vertices and
  links) are relative to the original graph,  red
  objects are formed by the decimation protocol. Solid links are
  part of the maximal tree, dashed links are
  chords. \label{fig:noprob}}
\end{figure}

\begin{enumerate}

\item {\bf new loops with no entropy production}:

The simplest case is realized when no chords connect any $j$ to
$0$ and no chords connect any $j$ to any $j'$, cf.\ Fig.~\ref{fig:noprob} for
one example. In this case, all original links
$j\to0$ ($A$, $B$, $C$ and $D$ in Fig.~\ref{fig:noprob}) are on the spanning tree and no
links join any $j$ to any $j'$. After the links and the central state
have been removed, the red links are created (a, b, c, d, e,
f in the example): they are in a number $n(n-1)/2$. A number
$(n^2-3n+2)/2>0$ (for any $n \ge 2$) are {\em new chords} (a, d and e
in the example), while the remaining $n-1$ are links of
the new spanning tree (b, c and f). Therefore new loops have been
created (in the example they are $3-4-2-3$ with chord a, $3-1-2-3$
with chord e and $1-2-4-1$ with chord d). It is immediate to
verify that the affinity of the new loops vanishes: for instance the
loop $3-1-2-3$ has forward transition rate given by $3\to 0, 0\to 1,
1\to 0, 0\to 2, 2\to 0, 0\to 3$ and backward transition rate given by
$3 \to 0, 0\to 2, 2\to 0, 0\to 1, 1\to 0, 0\to 3$ and they exactly
cancel out (the exit rates are omitted, but they cancel out trivially):
these new loops do not contribute to the entropy production.

\begin{figure}[ht]
\includegraphics[width=6cm,clip=true]{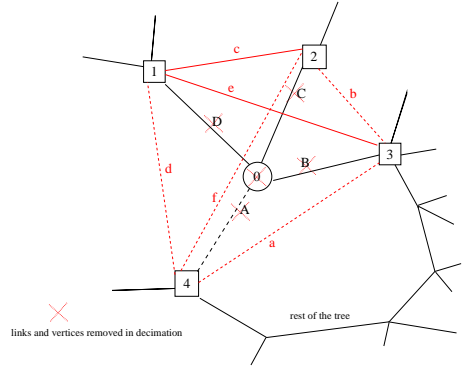}
\caption{An example of state-removal where a chord is removed by
decimation. Black objects (vertices and links) concern the
original graph, before decimation, red objects are the new ones formed
after decimation. Solid links are part of the maximal tree, dashed
links are chords. \label{fig:cordaint}}
\end{figure}

\item {\bf loop-shortening}:

Another possibility (see Fig.~\ref{fig:cordaint}) is that some link $0
\to j^*$ is a chord in the original graph, which means that it is not in
the spanning tree: e.g. link A in the example, with $j^*=4$. Then,
state $j^*$ is connected to 0 through some other unique path on
the tree, possibly passing through a state $j'$ (the unique path on
the tree is also represented in the figure, terminating with the
link $3-0$), forming a loop $0-j^*-tree-j'-0$. In this
case, the decimation of state $0$ creates the link $j^* \to
j'$ (link ``a'' in the example) as a chord of the
loop $j^*-tree-j'-j^*$, which is two steps shorter than the orginal
loop. It is immediate to see that the affinity of the new loop is the 
same as that of the old one. In this
case, all new links starting from $j^*$, or from $j'$, and ending in
another $j$, must be chords, since $j^*$ and $j'$ are joined by a
unique path on the tree which has not been touched by decimation:
the number of new chords is larger than in the previous case, but all 
their loops have zero affinities.

There is also the possibility that the loop passing through chord $A$
is simply given by $j^*-0-j-j^*$, i.e. that it is a
$3$-loop, originally belonging to the graph: this case can be put
in the last category, simply exchanging the roles of the chords A and a:
we call loop-crunching this case.

\begin{figure}[ht]
\includegraphics[width=6cm,clip=true]{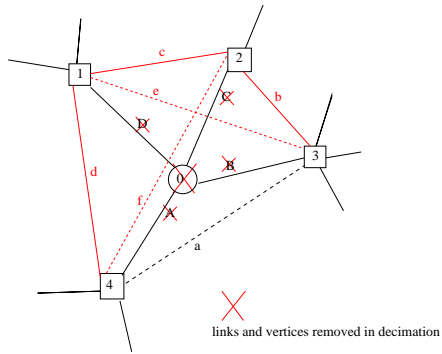}
\caption{An example of state-removal where a chord connects two states
  which are directly linked to the removed state. Black objects
  (vertices and links) pertain to the original graph, before
  decimation, red objects are formed by
  decimation. Solid links are part of the maximal tree, dashed links
  are chords.  \label{fig:cordaext}}
\end{figure}

\item {\bf loop-crunching}:

  The last possibility is that some link $j \to j'$ already existed in
  the original graph, which means that it is a chord of the loop $A-j-j'-A$,
cf.\ Fig.~\ref{fig:cordaext}, where chord a connects states 4 and 3. 
In this case the removal of
  state $A$ leads us to crunch the $3$-loop, making it a simple link
  $j-j'$ with a new transition rate. The original loop and its
  contribution to entropy production are then lost.

\end{enumerate}

We stress that a mathematical proof of the above considerations is still lacking,
although our arguments are strongly supported by numerical results.

%%%%%%%%%%%%%%%%%%%%%%%%%%%%%%%%%
\section{Concluding remarks and open problems}

In this paper we support, both numerically and theoretically, the idea
that fluctuations of the entropy production are essentially
insensitive to a coarse-graining based on decimation of fast states,
provided that decimation does not remove fundamental loops carrying
net currents. The threshold of coarse-graining level which
trigger the removal of such loops is not fully understood, but our
investigation suggests that entropy production fluctuations are
generally quite robust with respect to decimation.  Moreover, this
robustness does not appear directly related to the characteristic
times of the removed states. Robustness or fragility of loops appears
mostly related to the global structure of the network at hand.

We applied this analysis to the network model of the biomolecule known as
kinesin, discovering that no entropy production is lost if a coarse-graining 
from six to five states is performed. This observation is potentially interesting
in biophysics, since entropy production is a fundamental property of
irreversible chemical reactions, such as those fueling the kinesin motor
protein. On the contrary, the decimation of the model from six to four states 
is catastrophic, making the model unsuitable to produce work.

More detailed studies are necessary to quantify
the entropy production variations induced by coarse graining: the main missing
ingredient is the evaluation of the effect
of decimation on the currents of the surviving loops. This
will lead to a better understanding of the role and meaning of $\twt$ as
a definition of entropy production. In particular, understanding the relation
between $\twt$ and macroscopic observable properties of the system may help
in modelling non-equilibrium systems.

\begin{acknowledgments} 
  A. P. acknowledges the support of the ``Granular-Chaos'' project,
  funded by the Italian MIUR under the FIRB-IDEAS grant number
  RBID08Z9JE. S. P. wishes to thank M. Mueller for suggesting the
  kinesin example. L. R.  acknowledges the contribution of the European Research 
Council within the 7th Framework Programme (FP7) of the European Community (EC), 
ERC Grant Agreement n. 202680.  The EC is not responsible for any use that might 
be made of the data appearing herein.

\end{acknowledgments}

%%%%%%%%%%%%%%%%%%%%%%%%%%%%%%%%%
\appendix

%%%%%%%%
\section{The decimation procedure}

\label{ap:dec}

In this Appendix, we summarize the coarse graining method
introduced in~\cite{PV08}. Consider a master equation of the form of eq. 
(\ref{eq:master}). Due to the Markovian nature of the process, the time spent 
in a generic state $n$ is
exponentially distributed with average $\tau_n=1/W^0_n$. One may wish to
decimate all states having an average permanence time smaller than a
prescribed threshold $\Delta \tau$. To do that, Ref.\cite{PV08} sets to 0 
the time spent in these states. In this way, the fast states
disappear from the description and transitions to them are redirected
to other states with proper statistical weights. In formulae, if a
state $i$ is linked to a state $j$ via a fast state $n$ that must be 
eliminated, the transition rate $W_{i\to j}$ from $i$ to $j$ is renormalized
to yield the rate:
\begin{equation}\label{eq:decimate}
  \tW_{i\to j}=W_{i \to j}+W_{i \to n}W_{n \to j}/W^0_n
\end{equation}
If $W_{i \to j}=0$, the decimation creates a direct connection between the surviving
states, which is reminscent of the states that disappeared from the model under consideration.

This procedure corresponds to an adiabatic approximation and is commutative, if
the prescription of \cite{PV08} is followed. Once the set of states to be
decimated is determined by the threshold, they can be decimated in
any order without affecting the final result, as long as the set itself
is not modified during the decimation procedure. It may happen, indeed, 
that the permanence time of some of the states selected for decimation
becomes larger than $\Delta \tau$, while other states are decimated.
The recipe of \cite{PV08} requires that this state be eventually decimated
nonetheless.

\section{Effect of decimation on the current in a single loop}
\label{ap:singleloop}

In this Appendix, we investigate the effect of decimation on the
current of a single loop consisting of $N$ states.  For convenience, let us
rewrite the master equation:
$$
\frac{d}{dt} P_n(t)= W_{n-1\to n}P_{n-1}+W_{n+1\to n}P_{n+1}-P_n
(W_{n\to n+1}+W_{n\to n-1}) ~,
$$
with $P_0=P_N$, as:
\begin{equation}
\frac{d}{dt} P_n(t)= J_n-J_{n-1}
\end{equation}
where the {\em local} current $J_n$ is given by:
\begin{equation}
J_n=P_{n-1}W_{n-1\to n}-P_nW_{n\to n-1}.
\end{equation}
In a stationary state, the current is site-independent and one may write
$J_n=J$. In particular, detailed balance and equilibrium hold if $J=0$. 
The set of equation $J_n=J$, together with the normalization condition
$\sum_n P_n=1$, can be solved for both  $J$ and the invariant measure 
$P_n^{inv}$. For instance, let us proceed iteratively, as follows:
\begin{eqnarray}
P_n&=&P_{n-1} \frac{W_{n-1\to n}}{W_{n\to n-1}}-\frac{J}{W_{n\to n-1}}=\nonumber\\
&=&P_{n-2}\frac{W_{n-2\to n-1}W_{n-1\to n}}{W_{n\to n-1}W_{n-1\to n-2}}
-J\left(\frac{1}{W_{n\to n-1}}+\frac{1}{W_{n-1\to n-2}}\frac{W_{n-1\to n}}
{W_{n\to n-1}}
\right)=\dots\nonumber\\
&=&P_n\prod\limits_{k=1}^N \frac{W_{k-1\to k}}{W_{k\to k-1}}
-J\left(\sum\limits_{j=0}^{N-1}
\frac{1}{W_{n-j\to n-j-1}}\prod\limits_{k=0}^j\frac{W_{n-j+k-1\to n-j+k}}
{W_{n-j+k\to n-j+k-1}}\right).
\end{eqnarray}
We obtain $P_n$ from the last expression
\begin{equation}
P_n=\frac{-J\left(\sum\limits_{j=0}^{N-1}\frac{1}{W_{n-j\to n-j-1}}
\prod\limits_{k=0}^j
\frac{W_{n-j+k-1\to n-j+k}}{W_{n-j+k\to n-j+k-1}}\right)}
{1-\prod\limits_{k=1}^N \frac{W_{k-1\to k}}{W_{k\to k-1}}}
\end{equation}
and by means of the normalization condition $\sum P_n=1$,
we reach the following closed expression for $J$:
\begin{equation}\label{eqcurrent}
J=\frac{\left(\prod\limits_{k=1}^N \frac{W_{k-1\to k}}{W_{k\to k-1}}\right)-1}
{\sum\limits_{n=1}^N
\sum\limits_{j=0}^{N-1}\frac{1}{W_{n-j\to n-j-1}}\prod\limits_{k=0}^j
\frac{W_{n-j+k-1\to n-j+k}}{W_{n-j+k\to n-j+k-1}}}.
\end{equation}
Let us now decimate one fast state, $n^*$ say, 
and consider the current. It is easy to show that the
numerator is not affected by the decimation protocol defined by
eq. (\ref{eq:decimate}). Conversely, the denominator decreases by an
amount $\Delta D$ which can be espressed as follows:
\begin{equation}\label{eq:deltaD}
\Delta D= D_o-D_d=\frac{1}{W_{n^*\to n^*-1}}+
\sum\limits_{j=0}^{N-1}\frac{1}{W_{n^*+1\to n^*}}
\prod\limits_{k=0}^j\frac{W_{n^*+k\to n^*+k+1}}{W_{n^*+k+1\to n^*+k}}
\end{equation}
where $D_0$ and $D_d$ are the denominator in (\ref{eqcurrent}) for the
original and the decimated system, respectively. As $\Delta D$ is
positive, the current in the decimated system
is larger than in the original one,  the difference being
\begin{equation}
\Delta J=J_d-J=J \frac{\Delta D}{D_0-\Delta D}.
\end{equation}
This allows us to check what happens in simple
cases. For instance, eq. (\ref{eq:deltaD}) leads to $\Delta
D/D=1/N$, if all state have same left
and right jump rates (the two must be different to have a non-trivial
current). If the rate of the decimated state is much faster than the
others, eq. (\ref{eq:deltaD}) also shows that the correction
decreases linearly with the separation of time scales, i.e. with the ratio
of the average rates of the fast states and that of the other states. 
This is consistent with the picture of  
the current correction being essentially due to a rescaling of the times
related to the elimination of the fast state. In other words, the
magnitude of the correction seems to be always related to the ratio of
the time spent in the fast state(s)  and the time spent in the slow ones.

%%%%%%%%
\section{Graphs and currents}

\label{ap:graph}

We consider a Markovian (continuous time) process on $N$ states. The
$N$ states are considered as nodes of a graph. The transitions between
different states are considered as links (edges) between nodes.

%%%%%%%%%%%%%%%%%5
\subsection{Fundamental cycles}

Graph theory simplifies the classification of closed loops on a
graph~\cite{S76}, identifying a set of fundamental ``cycles''. Given a
graph $G$ with $N$ vertices (nodes) and $E$ edges (links between
nodes), the strategy - exemplified in Figure~\ref{fig:circuits} - is
the following:

\begin{figure}[ht]
\includegraphics[width=6cm,clip=true]{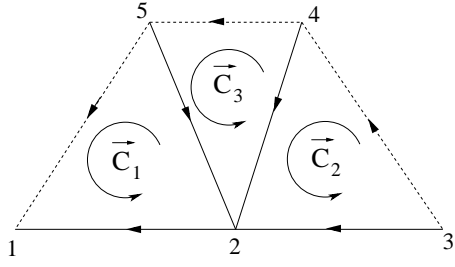}
\caption{An example of graph with $5$ states and three fundamental
loops: all transitions (links) have a given orientation. A possible
maximal tree is the one made of only solid links, with the three
dashed links representing the remaining chords, which individuate
three fundamental loops. Any other possible loop, e.g. $\vec{C}=1 \to
2 \to 4 \to 5 \to 1$, can be decomposed in the sum of fundamental
loops, e.g. $\vec{C}=\vec{C}_1+\vec{C}_3$. \label{fig:circuits}}
\end{figure}

\begin{itemize}

\item identify a maximal tree $T(G)$, i.e.\ a set containing all $N$
  vertices and part of the $E$ edges, which is connected and does not
  contain circuits. It is easy to show that $T(G)$ has $N-1$ edges.
  Many maximal trees can be identified, but one suffices;

\item given an arbitrary maximal tree $T(G)$, the edges of $G$ which
  do not belong to $T(G)$ are called chords of $T(G)$, they are
  in a number $\nu=E-N+1$; 

\item if only one chord $s_\alpha$, $\alpha \in [1,\nu]$, is added
  to $T(G)$, the new graph contains only one circuit,
  $C_\alpha$, obtained by $T(G)+s_\alpha$ removing all edges which are
  not part of the circuit; therefore, from a maximal tree, $\nu$
  circuits can be generated adding the $\nu$ chords; 

\item the set of $\nu$ circuits obtained from the $\nu$ chords of a
  maximal graph is called a fundamental set of circuits, 
  denoted by $\{C_1,...,C_\nu \}$

\item  orientation of edges must be introduced: each edge is
  assumed to be oriented in an arbitrary direction, giving the
  oriented version of $G$, denoted as $\vec{G}$; then one can take a
  subgraph with oriented edges $\vec{P}$, which may have
  different orientations with respect to the original orientations
  of the edges of $\vec{G}$. The function $S_e(\vec{P})$ is introduced for
  these cases: it returns $1$ if the edge $e$ is in $\vec{P}$ and has
  the original orientation, $-1$ if it is in $\vec{P}$ and has
  opposite orientation, and $0$ if $e$ is not in $\vec{P}$.

\item a cycle is an oriented circuit, e.g. $\vec{C}$; a fundamental
cycle is denoted by $\vec{C}_\alpha$: for simplicity we always choose
the orientation of a fundamental circuit to be parallel to the
orientation of its chord $\alpha$, i.e. $S_\alpha(\vec{C}_\alpha)=1$;

\item the scalar product among cycles is defined as
\begin{equation}
(\vec{C},\vec{C}_\alpha)=S_\alpha(\vec{C})S_\alpha(\vec{C}_\alpha)\equiv S_\alpha(\vec{C})
\end{equation}
where $\alpha$ is the chord which generates the circuit $C_\alpha$;
this scalar product can only take three values: $0$, $1$ or $-1$.

\item a decomposition of cycles is finally achieved: any cycle
  (oriented circuit) of the graph $G$ can be linearly decomposed using
  the fundamental set as a basis:
\begin{equation}
\vec{C}=\sum_{\alpha=1}^\nu(\vec{C},\vec{C}_\alpha)\vec{C}_\alpha
\end{equation}

\end{itemize}

%%%%%%%%%%%%%%%%%%%%
\subsection{Currents}

The current for the $\omega \to \omega'$ transition is
\begin{equation} \label{cur_def}
J(\omega \to \omega',t)=P_\omega(t)W_{\omega \to \omega'}-P_{\omega'}(t)W_{\omega' \to \omega}.
\end{equation}
The stationary state value is denoted by $J(\omega \to \omega')$.
The stationarity condition $dP_\omega^{inv}/dt=0$ is equivalent to
\begin{equation}
\sum_{\omega'} J(\omega' \to \omega)=0 \;\;\;\;\;\;\; \forall \omega
\end{equation}
which is known as {\em Kirchhoff current law}. If the transition $\omega
\to \omega'$ corresponds to the oriented edge $e$, its steady state current is
also denoted as $J_e$.

The current (or flux) on a fundamental circuit is defined as the
steady state transtion current flowing in the chord $\alpha$ in the 
original direction and is denoted by $J_\alpha$. For instance if $\alpha$ 
is the oriented edge corresponding to the transition $\omega \to \omega'$, 
then $J_\alpha=J(\omega \to \omega')$ and the flux of the associated cycle
$\vec{C}_\alpha$ is equal to $+J_\alpha$.

The Kirchhoff law for the steady state guarantees that a current on any
edge is the sum of the currents going through the cycles which
intersect the edge, i.e.
\begin{equation} \label{edgecur}
J_e=\sum_{\alpha=1}^\nu S_e(\vec{C}_\alpha)J_\alpha.
\end{equation}
An edge of the graph $G$
can be oriented in a different direction with respect to the edges of
the cycles, therefore the sign function $S_e$ is used.

The fluctuating instantaneous current $J_\alpha$ depends on the particular
realization of the Markov process; it is measured on a chord $\alpha$ as:
\begin{equation}
j_\alpha(t)=\sum_{n=-\infty}^{+\infty}S_\alpha(e_n)\delta(t-t_n)
\end{equation}
where $t_n$ is the time of the random transition $e_n$ (an oriented edge 
of the graph) during a trajectory of the stochastic
process. In brief, $j_\alpha$ is the instantaneous and oriented
rate of the transitions in the chord $\alpha$, for
a particular realization of the process. It is a stochastic variable.  Its
time-average (in a finite time $t$) is denoted by
\begin{equation}
G_\alpha(t)=\frac{1}{t}\int_0^t dt' j_\alpha(t'),
\end{equation}
which is still a stochastic variable.
Some properties of $j_\alpha$ and $G_\alpha$ have been studied in~\cite{AG07}.

%%%%%%%%%%%%%%%%%%%%
\subsection{Affinities}

The affinity of a transition $\omega \to \omega'$ is defined as
\begin{equation}
A(\omega \to \omega',t)=\ln \frac{P_\omega(t)W_{\omega \to
\omega'}}{P_\omega'(t)W_{\omega' \to \omega}}
\end{equation}
The affinity of a cycle $C$ is defined as
$A(C)=\sum_e S_e(C)A(e)$, but it can also be defined as
$A(C)=\sum_e S_e(C)B(e)$.
where 
$B(\omega \to \omega')=\ln (W_{\omega \to \omega'}/W_{\omega' \to
\omega})$.
The equivalence of these two
forms is due to the fact that all $P_\omega(t)$
cancel out, in a cycle. For this reason the affinity of a cycle does not depend
upon time, but only on the transition rates, which come from the ``external
physical constraints'', e.g. mechanical, chemical and thermodynamical
forces. 

Thanks to the decomposition of cycles described above,
one can linearly decompose the affinity of any cycle in terms of
affinities of a ``fundamental set of cycles'' $\{\vec{C}_\alpha\}$:
\begin{equation}
A(\vec{C})=\sum_\alpha (\vec{C},\vec{C}_\alpha) A(\vec{C}_\alpha)
\end{equation}
where $(.,.)$ is the previously defined scalar product between cycles.

%%%%%%%%%%%%%%%%%%%%
\subsection{Entropy production}

Having defined the Gibbs entropy as
\begin{equation}
\ent(t)=-\sum_\omega P_\omega(t) \ln P_\omega(t),
\end{equation}
its time derivative can be decomposed in two parts
$d\ent/dt=d_e \ent/dt+d_i \ent/dt$,
where the bilinear form
\begin{equation}
\frac{d_i \ent}{dt}=\frac{1}{2}\sum_{\omega,\omega'}J(\omega \to \omega',t)A(\omega \to \omega',t) \ge 0
\end{equation}
is considered as the internal entropy production, and the rest $d_e \ent/dt$
is the entropy flux through the boundaries of the system of interest. 
In the steady state one has
$d_e \ent/dt=-d_i \ent/dt$.

A definition of entropy production per trajectory is given by Lebowitz
and Spohn~\cite{LS99}, see Eq.~\eqref{ls_def}.  It depends on a
particular realization, i.e. it is a stochastic variable. It can also
be written as:
\begin{equation} \label{ls2}
\twt=\frac{1}{t}\sum_eB(e)\int_0^tdt'j_e(t').
\end{equation}

Lebowitz and Spohn have noticed that
\begin{equation}
\lim_{t \to \infty} \langle \twt \rangle=\left.\frac{d_i \ent}{dt}\right|_{st}
\end{equation}
in the stationary state. The following relation has instead been noticed in Ref.\cite{AG07}:
\begin{equation} \label{res}
\twt=Q_t+R_t
\end{equation}
with
\begin{align}
Q_t&=\sum_\alpha A(\vec{C}_\alpha)G_\alpha(t)\\
R_t&=\frac{1}{t}\sum_{e \neq {\alpha}} B(e)\left[\int_0^t dt' \left(j_e(t')-
\sum_\alpha S_e(\vec{C}_\alpha)j_\alpha(t')\right) \right].
\end{align}
The $Q_t$ term is the contribution due to the fundamental set of cycles.
The ``remainder'' $R_t$ has zero average (thanks to the Kirchhoff law
Eq.~\eqref{edgecur}). This implies that
\begin{equation}
\left.\frac{d_i \ent}{dt}\right|_{st}=\lim_{t \to \infty} \langle \twt \rangle=
\lim_{t \to \infty} \langle Q_t \rangle=\sum_\alpha A(\vec{C}_\alpha)J_\alpha,
\end{equation}
since
\begin{equation}
\lim_{t \to \infty}\langle G_\alpha(t) \rangle=\lim_{t \to \infty} \frac{1}{t}\left\langle \int_0^t j_\alpha(t') dt' \right\rangle=J_\alpha.
\end{equation}
Numerical comparison of the fluctuations of $Q_t$ and those of $\twt$
show that they have identical Cramer functions (see
Figure~\ref{fig:equiv}), in many examples of continuous
time Markov processes.

From eq.~\eqref{cur_def}, detailed balance, with respect to the invariant measure, is equivalent to
\begin{equation}
J(\omega\to\omega')=0 \;\;\;\;\; \forall (\omega \to \omega'),
\end{equation}
which implies that the probability of any trajectory is equal to the
probability of its time reversal.  
Detailed balance also implies that the flux on
any cycle vanishes, $J_\alpha=0$,
and that affinities vanish on a single edge as well as on any cycle, eg.
$A(\vec{C}_\alpha)=0$.
As an immediate consequence, the internal entropy production vanishes:
\begin{equation}
\left.\frac{d_i \ent}{dt}\right|_{st}=0.
\end{equation}

%%%%%%%%
\section{Appendix: parameters in the kinesin model}
\label{app_kin}

As sketched in Section~\ref{sec:kin}, the rates in the kinesin network model of
Ref.\cite{LL07} are adjusted to the parameters obtained by
specific experiments. Moreover, they depend on the concentrations of
the chemical species entering the reaction (ADP, ATP and P), as well
as on the load force $F$. More formally, one has:
\begin{equation}\label{kin_rate_def}
W_{i \to j}=k_{ij}\ I_{ij}([X])\ \Phi_{ij}(F)
\end{equation}
where the $k$'s are the experiment-specific parameters. The functions
$I_{ij}$ and $\Phi_{ij}$ express the dependence of the reaction rates on a
generic chemical species $X$ and/or on the load force $F$. If the
transition from $i$ to $j$ does not involve chemical binding, we define
$I_{ij}\equiv 1$.

Assuming diluted solutions, all reactions are diffusion-limited, so
that we can assume $I([X])\sim [X]$. The $\Phi$'s are adimensional
functions, with the convention $\Phi(0)=1$. Theoretical considerations
lead to $\Phi_{ij}(F)=\Phi_{ji}(F)=2/(1+e^{\chi_{ij}\bar{F}})$ for the
chemical transitions, i.e. all but those between states $2$ and $5$.
Mechanical transitions are parametrized by $\Phi_{25}=e^{-\theta
  \bar{F}}$ and $\Phi_{52}=e^{(1-\theta) \bar{F}}$. The $\chi$'s and
$\theta$ are additional parameters obtained by experiments, while
$\bar{F}=lF/kT$ is the adimensional force ($l\approx 8 nm$ being the
average kinesin step length and $k$ the Boltzmann constant). With these
choices, the $k$'s are dimensionally different depending on whether
they multiply a concentration (dimensions of rate divided by
concentration, $[(\mu M\ s)^{-1}]$) or not (dimension of a rate,
$[s^{-1}]$).

The parameters we used in the simulations are derived from those
reproducing the results of the experiment \cite{CC05}:

\begin{itemize}
\item The values of $k$'s describing the experiment \cite{CC05}
  according to \cite{LL07} are $k_{25}=3\cdot10^5$,
  $k_{52}=0.24$, $k_{12}=k_{45}=2.0$, $k_{21}=100$,
  $k_{56}=k_{61}=k_{23}=k_{34}=100$, $k_{65}=k_{32}=0.02$,
  $k_{16}=k_{43}=0.02$. The upper and lower cycle in
  Fig. \ref{kinesin_scheme} are assumed to have same parameters, apart
  from the transition from $5$ to $4$, which is determined from
  theoretical considerations as $k_{54}=k_{21}(k_{52}/k_{25})^2=6.4 \
  10^{-11}$.
\item Typical concentrations in the experiment are $0.5\mu M$. For
  simplicity, we assume all of them to be kept constant and equal to
  $[P]=[ADP]=[ATP]=1\mu M$.
\item The mechanical parameters reproducing the results of experiment
  \cite{CC05} are: $\theta=0.65$, $\chi_{12}=\chi_{45}=0.25$,
  $\chi_{23}=\chi_{56}=0.15$, $\chi_{34}=\chi_{61}=0.15$. In all
  cases, we have $\chi_{ij}=\chi_{ji}$.
\end{itemize}

In section~\ref{sec:kin}, we considered two instances of the model. The first one
is without load, $F=0$. In this case and with the assumptions above,
it is easy to obtain the transition rates: all the $\Phi$'s and
concentrations are equal to $1$, so from Eq. (\ref{kin_rate_def}) we
obtain $W_{i\to j}=k_{ij}$: the rates are just the $k$'s listed above.

About the load case, the unit of the adimensional force is equal to
$kT/l\approx 0.5 pN$. Experiments are performed with forces of the
order of piconewton. We took a value $\bar{F}=10$: substituting this value
in the expression for the $\Phi$'s leads to the following values
of the transition rates, which are those used in the simulations of
the model with load: $W_{2\to 5}=451$, $W_{5\to 2}=7.95$,
$W_{1\to 2}=W_{4\to 5}=0.3$, $W_{2\to 1}=15$,
$W_{5\to 6}=W_{6\to 1}=W_{2\to 3}=W_{3\to 4}=36.5$,
$W_{6\to 5}=W_{3\to 2}=0.007$, $W_{1\to 6}=W_{4\to 3}=0.007$,
$W_{5\to 4}=0.5\ 10^{-11}$.

\section{List of the main symbols}

\begin{itemize}

\item  $W_{i \to j}$ is the transition rates from $i$ to $j$

\item  $W_i^0$ is the exit rate from state $i$

\item $\tau_i=1/W_i^0$ is the characteristic time of state $i$

\item $P_n(t)$ is the probability of being in $n$ at time $t$

\item $P_n^{inv}$ is the invariant probability of being in $n$ 

\item $\Delta \tau$ is the time threshold for decimation

\item $\tW_{i \to j}$ are the new transition rates in the decimated process

\item $\twt$ is the Lebowitz-Spohn entropy production integrated on time $t$ and divided by $t$

\item $C()$ is the Cramer's function of the entropy production

\item $f(\twt) \sim e^{-tC(\twt)}$ is the probability density of $\twt$

\item $C_{\Delta \tau}(\twt)$ is the Cramer function in the decimated process with a time threshold $\Delta \tau$.

\item $\vec{C}_\alpha$ is an oriented cycle of the graph

\item $G_\alpha(t)$ is the current on cycle $\alpha$ averaged on a finite time $t$

\item $A(\vec{C}_\alpha)$ is the affinity associated to the oriented cycle $\vec{C}_\alpha$.

\end{itemize}

%%%%%%%%%%%%%%%%%%%%%%%%%%%%%%%%%%%%
%%%%%%%%%%%%%%%%%%%%%%%%%%%%%%%%%%%

\bibliography{fluct.bib}

\end{document}